\documentclass[11pt]{article}
\usepackage[dvips]{graphicx}
\usepackage{amsmath,latexsym,amssymb,epsfig,psfrag}
\usepackage{pdproc}

\newcommand{\be}{\begin{equation}}
\newcommand{\ee}{\end{equation}}

\newcommand{\de}{\partial}

  \makeatletter 
  \def\@cite#1{[#1]} 
  \makeatother    
\begin{document}

\renewcommand{\thefootnote}{\alph{footnote}}

\title{
Supersymmetry Breaking and Restoration in the Interval
~\footnote{Talk based on Ref.{\cite{us}}}}

\author{Ver\'onica Sanz}

\address{ 
IPPP Physics Department, Durham University \\
DH1 3LE (Durham) UK \\ 
and \\
 Departament de F\'\i sica Te\` orica, IFIC, Universitat de Val\` encia-CSIC \\
E-46071 Val\` encia, Spain
\\ {\rm E-mail: veronica.sanz@ific.uv.es}}

\abstract{
We study fermions, such as gravitinos and gauginos in supersymmetric
theories, propagating in a five-dimensional bulk where the fifth
dimension is  an interval. We show the mass spectrum becomes independent
from the Scherk-Schwarz parameter  if the boundary mass terms obey a relation of alignment with the bulk supersymmetry breaking.
}

\normalsize\baselineskip=15pt

\section{The method: Fermions in the interval}

 In a manifold ${\cal M}$ with a boundary the dynamics is
determined by two equally important ingredients: the bulk equations of
motion and the boundary conditions (BC's). An economical way to
determine a set of consistent BC's together with the bulk equations of
motion is the action principle~\footnote{For an alternative approach
see~\cite{csfermion}.}: under a variation of the dynamical fields the
action must be stationary.

Since we are mainly interested in supersymmetric theories, we will
take the fermions to be symplectic-Majorana spinors.  In particular we will consider the
gaugino case, the treatment of gravitinos being completely analogous.
The 5D spinors $\Psi^i$ satisfy the symplectic-Majorana reality
condition and we can represent them in terms of two chiral 4D spinors
according to~\footnote{We use the Wess-Bagger convention~\cite{wb} for
the contraction of spinor indices.}
\begin{equation}\label{uno}
\Psi^i = \begin{pmatrix} \eta^i_\alpha \\ 
\bar{\chi}^{i\,\dot{\alpha}} \end{pmatrix} \, , \qquad
\bar{\chi}^{i \,\dot{\alpha}} \equiv \epsilon^{ij} \, 
\left(\eta^j_\beta \right)^\ast \,  \epsilon^{\dot{\alpha} 
\dot{\beta}}  
\; .
\end{equation}
 Consider thus the bulk Lagrangian
\begin{equation} 
{\cal L}_{\text{bulk}} =i\, \bar{\Psi} \gamma^{{}_M}
D_{{}_M} \Psi =\frac{i}{2} \bar{\Psi} \gamma^{{}_M}
D_{{}_M} \Psi - \frac{i}{2} D_{{}_M} \bar{\Psi} \gamma^{{}_M} \Psi \; .
\end{equation}
where the last equation is not due to partial integration but holds
because of the symplectic-Majorana property, Eq.~(\ref{uno}). The
derivative is covariant with respect to the $SU(2)_R$ automorphism
symmetry and thus contains the auxiliary gauge connection $V_M$.  The
field $V_M$ is non propagating and appears in the off-shell
formulation of 5D supergravity~\cite{zucker}.  A vacuum expectation
value (VEV)~\footnote{Consistent with the bulk equation of motion
  $d\,(\vec q\cdot \vec V)=0$~\cite{zucker}.}
\be V_{_M} = \delta^5_{_M} \, \frac{\omega}{R} \,
\vec{q}\cdot\vec\sigma \, , \qquad {\vec{q}}^{\, 2}=1
\label{SSVEV}
\ee
implements a Scherk-Schwarz (SS) supersymmetry breaking
mechanism~\cite{ss} in the Hosotani basis~\cite{hos,sshos}.  The unitary vector
$\vec{q}$ points toward the direction of SS breaking.  We supplement
the bulk action by the following boundary terms at $y=y_f$ ($f=0,\pi$)
with $y_0=0$ and $y_\pi=\pi R$
\begin{equation} 
{\cal L}_{f} = \frac{1}{2} \bar{\Psi} \left(T^{{(f)}} + \gamma^5 \,
V^{{(f)}} \right) \Psi =\frac{1}{2} \,{\eta^i}M^{_{(f)}}_{ij} \eta^j
+{\rm h.c.}\; , 
\label{bmasses}
\end{equation}
The variation of the bulk action gives 
\begin{equation}
\delta S_{\text{bulk}} = \int d^5x \, i \left(\delta \bar{\Psi}
\gamma^{{}_M} D_{{}_M} \Psi - D_{{}_M} \bar{\Psi} \gamma^{{}_M}
\delta\Psi \right) - \int d^4x \, \left[\delta \eta^i \,\epsilon_{ij} \eta^j 
 + h.c. \right]^{\pi R}_0 \; ,
\label{bvar}
\end{equation}
where the boundary piece comes from partial integration.  One now has
to add the variation of the boundary action.  Enforcing that the total
action $S= S_{\text{bulk}}+ S_{\text{boundary}}$ has zero variation we
get the standard Dirac equation in the bulk provided that all the
boundary pieces vanish. The latter are given by
\begin{equation}
\left.\left[\delta \eta^i \left( \epsilon_{ij} +
M_{ij}^{{}_{(f)}}\right) \eta^j + \, \text{h.c.}  \right]\right|_{y
= y_f} = 0 \; . \label{bf}
\end{equation}
Since we are considering unconstrained variations of the fields, the
BC's we obtain from Eqs.~(\ref{bf}) are given by
\begin{equation}
\left.\left( \epsilon_{ij} + M_{ij}^{{}_{(f)}}\right) \eta^j
\right|_{y = y_f} = 0 \; . \label{bf2}
\end{equation}
These equations only have trivial solutions (are overconstrained)
unless
\be
\det\left(\epsilon_{ij} + M_{ij}^{{}_{(0)}}
 \right) =\det\left( \epsilon_{ij} + M_{ij}^{{}_{(\pi)}}\right)=0\;.
\label{det}
\ee
Imposing these conditions, we get the two complex BC's which are
needed for a system of two first order equations.  Note that this
means that an arbitrary brane mass matrix does not yield viable BC's;
in particular a vanishing brane action is inconsistent~\footnote{In
the sense that the action principle does not provide a consistent set
of BC's as boundary equations of motion.} since
$\det(\epsilon_{ij})\neq 0$~\footnote{Notice that this agrees with the
methods recently used in Ref.~\cite{moss}.}. 

The BC's resulting from Eqs.~(\ref{bf2}) are of the form
\begin{equation}
\left.\left(\eta^2 -z_f \, \eta^1 \right) \right|_{y =y_f} = 0,\quad z_f=-\frac{M_{11}^{_{(f)}}}{1+M_{12}^{_{(f)}}}
 =\frac{1-M_{12}^{_{(f)}}}{M_{22}^{_{(f)}}}
 \quad .
\label{zBC}
\end{equation}

The mass spectrum is found by solving the EOM with the boundary
conditions (\ref{zBC}).  To simplify the bulk
equations of motion it is convenient to go from the Hosotani basis
$\Psi^i$ to the SS one $\Phi^i$, related by the transformation
\begin{equation} 
\Psi = U \, \Phi ,\quad U=
\exp{\left(-i\, \vec{q} \cdot \vec{\sigma}\, \omega\, \frac{y}{R}\right)}  \;  \Rightarrow  i \, \gamma^{{}_M} \de_{{}_M} \Phi = 0 \quad . 
\label{cb}
\end{equation}
We now decompose the chiral spinor $\eta^i(x,y)$ in the Hosotani basis
as $\eta^i(x,y) = \varphi^i(y) \psi(x)$, with $\psi(x)$ a 4D chiral
spinor.
As a consequence of the transformation (\ref{cb}) the SS parameter
$\omega$ manifests itself only in the BC at $y=\pi R$~\footnote{Notice
that $U(y=0)=1$.  The roles of the branes and hence of $z_\pi $ and
$z_0$ can be interchanged by considering the SS transformation
$U'(y)\equiv U(y-\pi R)$.}:
\be \zeta_0\equiv\left.\frac{\phi^2}{\phi^1}\right|_{y=0}=z_0,\quad
\zeta_\pi\equiv\left.\frac{\phi^2}{\phi^1}\right|_{y=\pi R}=\frac{
\tan(\pi\omega) (i q_1 - q_2 - i q_3\, z_\pi) + z_\pi } {\tan(\pi\omega) (i q_1
\, z_\pi + q_2\, z_\pi + i q_3) + 1 }\;,
\label{SSbc}
\ee 
where $\zeta_f$ are the BC's in the SS basis. In particular the
boundary condition $\zeta_\pi$ is a function of $\omega$, $\vec q$ and
$z_\pi$.  From this it follows that we can always gauge away the SS
parameter $\omega$ in the bulk Lagrangian going into the SS basis
through (\ref{cb}). However now in the new basis $\omega$ reappears in
one of the BC's.

The bulk equations have the following generic solution
\begin{equation}
 \phi(y)= 
  \begin{pmatrix}     
    \bar a \cos(my)+\bar z_0 a \sin(my)
    \\
    -a\sin(my)+z_0 \bar a \cos(my)
  \end{pmatrix} º , \quad  a=\frac{z_0-\zeta_\pi}{|z_0-\zeta_\pi|}+\frac{1+z_0 \bar
  \zeta_\pi}{|1+z_0\bar \zeta_\pi |}\; .
 \label{sol}
\end{equation}
The solution (\ref{sol}) satisfies the BC's Eq.~(\ref{SSbc}) for the
following mass eigenvalues
\begin{equation}
m_n  = \frac{n}{R} + \frac{1}{ \pi
R }\,\arctan\left|\frac{z_0-\zeta_\pi}{1+z_0\,\bar \zeta_\pi}\right|\; .
\label{masses}
\end{equation}

\section{Supersymmetry breaking: independence on the Scherk-Schwarz breaking scale}

Physically inequivalent BC's span a complex projective space
$\mathbb{C}P^1$ homeomorphic to the Riemann sphere.  In particular,
$z_f =0$ leads to a Dirichlet BC for $\eta_2$, and the point at
infinity $ z_f =\infty$ leads to a Dirichlet BC for $\eta_1$.  Notice
that these BC's come from $SU(2)_R$ breaking mass terms. Special
values of $z_f$ correspond to cases when these terms preserve part of
the symmetry of the original bulk Lagrangian. In particular when both
the SS and the preserved symmetry are aligned those cases can lead to
a {\it persistent} supersymmetry as we will see. Generically,

\begin{center} 
{\it  when $z_0=\zeta_\pi$ there is a zero mode and
supersymmetry remains unbroken.}
\end{center}

{\it i.)} When the only sources of supersymmetry breaking
reside on the branes, setting them to cancel each other, $z_0=z_{\pi}$,
preserves  supersymmetry~\cite{horava}. 

{\it ii.)} Once supersymmetry is further 
broken in the bulk, an obvious way to restore it is by
determining $z_\pi$ as a function of $z_0$ and $\omega$ using the
relation (\ref{SSbc}) with $\zeta_\pi=z_0$.  This will lead to an
$\omega$-dependent brane-Lagrangian at $y=\pi R$. In this case we
could say that supersymmetry, that was broken by BC's (SS twist) is
{\it restored} by the given SS twist (BC's)~\cite{SSBC}.

{\it iii.)} There is however a more interesting case: suppose the brane Lagrangian
determines $z_\pi$ to be
\begin{equation}
z_\pi=z(\vec q\,)\equiv\frac{\lambda-  q_3}{q_1 - i q_2}
\label{vH}.
\end{equation}
with $\lambda=\pm 1$.  This special value of $z_\pi$ is a fixed point
of the SS transformation, i.e.~$\zeta_f=z_f$.  For $z_\pi=z(\vec q\,)$
the spectrum becomes independent on $\omega$.  In other words, for
this special subset of boundary Lagrangians, the VEV for the field
$\vec q\cdot\vec V_5$ does not influence the spectrum. The reason for
this can be understood by going back to the Lagrangian which we used
to derive the BC's. From the relation (\ref{zBC}) one can see that
condition (\ref{vH}) is satisfied by the mass matrix
\begin{eqnarray}
M_{12}^{_{(\pi)}} &=& \lambda q_3 \nonumber \\
M_{11}^{_{(\pi)}} &=& -\lambda (q_1 + i q_2 ) \nonumber \\
M_{22}^{_{(\pi)}} &=& \lambda(q_1 -  i q_2 ) 
\end{eqnarray} 
which can be translated into a mass term at the boundary $y=y_\pi$
along the direction of the SS term, i.e.~$V^{(\pi)}=0$ and $T^{(\pi)}=
- \lambda\,\vec q\cdot\vec\sigma$ in the notation of
Eq.~(\ref{bmasses}). In particular this brane mass term preserves a
residual $U(1)_R$ aligned along the SS direction $\vec{q}$.  In other
words, the SS-transformation $U$ leaves both brane Lagrangians
invariant and $\omega$ can be gauged away.  When we further impose
$z_0=z(\pm\vec q\,)$, i.e.~$V^{(0)}=0$ and $T^{(0)}=\pm T^{(\pi)}$ the
$U(1)_R$ symmetry is preserved by the bulk. In particular if
$z_0=z(\vec q\,)$ supersymmetry remains unbroken, although the VEV of
$\vec q\cdot \vec V_5$ is nonzero. One could say that in this case the
theory is {\it persistently} supersymmetric even in the presence of
the SS twist, with mass spectrum $m_n=n/R$. On the other hand if
$z_0=z(-\vec q\,)$ the theory is ({\it persistently})
non-supersymmetric and independent on the SS twist: the mass spectrum
is given by $m_n=(n+1/2)/R$. In this case supersymmetry breaking
amounts to an extra $\mathbb Z_2^\prime$
orbifolding~\cite{Barbieri:2000vh}.

Something similar happens in the warped case: when bulk
cosmological constant and brane tensions are turned on, invariance of
the action under local supersymmetry requires gravitino mass terms on
the brane. In the tuned case, -- i.e.~in the Randall-Sundrum (RS)
model -- those brane mass terms precisely give rise to the BC
$z_0=z_\pi=z(\vec q)$~\cite{bagger1}.  Note that there $\vec
q\cdot\vec V_5$ is replaced by $A_5$, the fifth component of the
graviphoton.  In fact, it has been shown that in this case there
always exists a Killing spinor and supersymmetry remains
unbroken~\cite{bagger2,lalak}, consistent with the result that in RS
supersymmetry can not be spontaneously broken~\footnote{A discrete
supersymmetry breaking by BC's, $z_0=z(-\vec q)$, $z_\pi=z(\vec q\,)$,
was performed in Ref.~\cite{Gherghetta:2000kr}.}  by the SS
mechanism~\cite{Hall:2003yc,bagger1}.


\section{Acknowledgements}

\noindent I would like to thank  T.~Okui  for
useful discussions and to the Particle Theory Department at Boston
University. I would also like to thank
the Theory Department of IFAE, where part of this work has been
done, for hospitality.

\bibliographystyle{plain}

\end{document}